\documentclass[cameraready]{Interspeech}

\usepackage{amsmath,graphicx,hyperref}
\usepackage[table]{xcolor}
\usepackage{algorithm}
\usepackage{algorithmic}
\usepackage{subcaption}
\usepackage{amssymb}
\usepackage{multirow,booktabs}
\usepackage{tabularx}

\title{ContextCodec: Content-Focused Context Guidance for Ultra-Low Bitrate Speech Coding}

\author[affiliation={1}]{Chengbin}{Liang}
\author[affiliation={2},correspondingauthor=true]{Wenqi}{Guo}
\author[affiliation={1}]{Hao}{Cao}
\author[affiliation={1}]{Zhijin}{Qin}

\address{
    $^1$ Department of Electronic Engineering, Tsinghua University, Beijing, China \\
    $^2$ Department of Automation, Tsinghua University, Beijing, China
}

\email{lcb25@mails.tsinghua.edu.cn}

\keywords{Neural speech codec, Ultra-low bitrate, CLIP-style phoneme alignment, Autoregressive latent refinement}

\begin{document}

\maketitle

\begin{abstract}
Neural speech codecs enable low-bitrate speech communication, yet at ultra-low bitrates ($< 1000$ bps) preserving perceptual quality and intelligibility is challenging. Existing designs often prioritize acoustic details, leaving limited capacity for the core linguistic message under tight bitrate constraints. To address this, we propose ContextCodec, a codec that transmits content-focused context features to explicitly guide reconstruction. ContextCodec adopts a dual-branch encoder that decouples acoustic details from content-focused context. The context branch is trained with a CLIP-style contrastive loss that aligns context features with phoneme indices, reducing paralinguistic leakage. During decoding, these features are injected at each decoding stage for explicit guidance. In addition, we introduce a lightweight autoregressive latent refinement module. Experiments show a strong quality-intelligibility trade-off down to 500 bps, with an RTF of 0.4886 on a typical mobile CPU.

\end{abstract}

\section{Introduction}
\label{sec:introduction}

Neural speech codecs (NSC) aim to convert continuous speech signals into compact discrete representations while preserving perceptual quality and intelligibility. They are widely used in audio tokenization \cite{borsos2023audiolm,zheng2025freecodec,Qwen3-TTS} and telecommunications applications \cite{lyra,lpcnet}. As speech communication is increasingly deployed in bandwidth-constrained environments such as satellite links, ultra-low-bitrate speech coding has become increasingly necessary \cite{lpcnet,zhang2025lspnet}.  However, once the bitrate is pushed to the ultra-low regime, speech coding becomes a zero-sum bit-allocation problem: bits used to preserve how speech sounds inevitably reduce the capacity available to convey what is being said. As a result, maintaining both perceptual naturalness and intelligibility becomes fundamentally difficult.

\begin{figure}[t]
    \centering
    \includegraphics[width=\columnwidth]{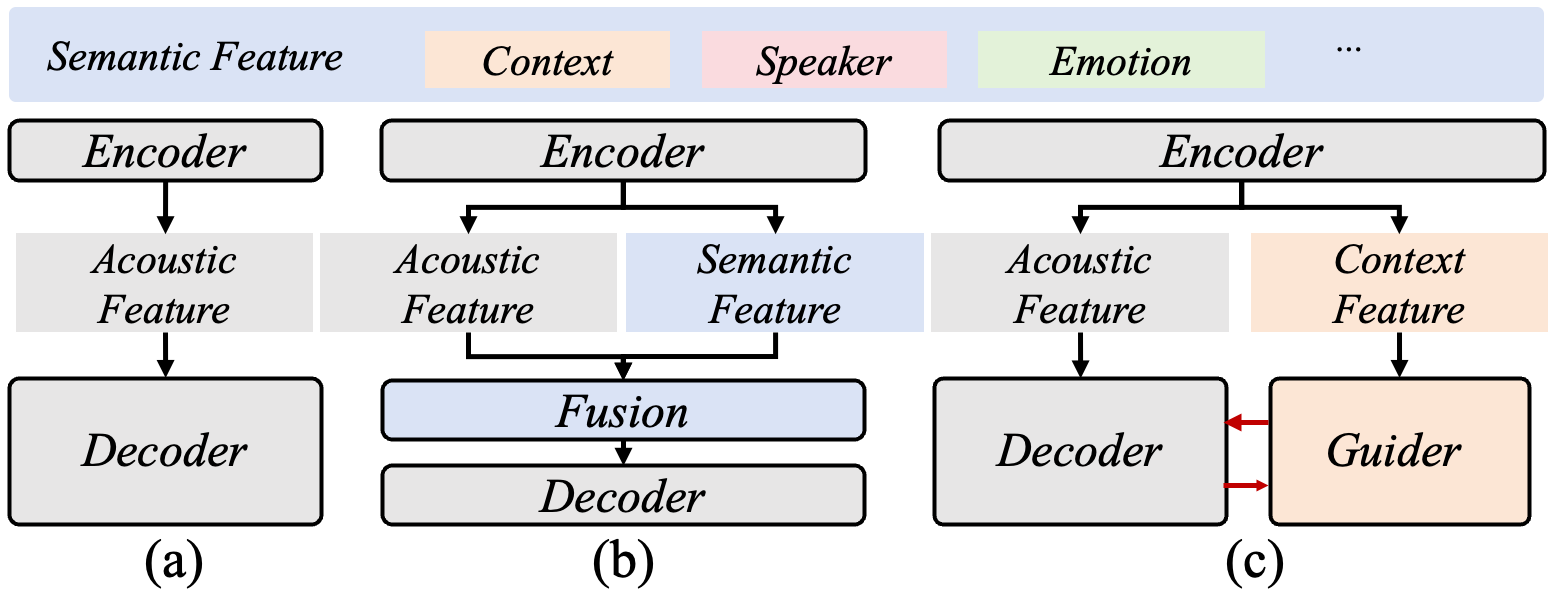}
    \caption{Different speech codec architectures. (a) Neural acoustic codecs. (b) Hybrid codecs. (c) ContextCodec(ours).}
    \label{fig:introduction}
    \vspace{-20pt}
\end{figure}

Existing neural speech codecs can be broadly grouped into two families, as shown in Figure~\ref{fig:introduction}(a-b). Neural acoustic codecs \cite{soundstream} reconstruct speech waveforms through adversarial training\cite{goodfellow2014generative,zheng2024srcodec}, with improvements in quantizer design \cite{dac,siuzdak2024snac,hificodec}, discriminator architectures \cite{encodec}, and structure \cite{xu2024lightcodec}. While these methods achieve strong perceptual quality, their objectives are still dominated by acoustic fidelity, encouraging to preserve speaker timbre, or other fine acoustic details. Under an ultra-low-rate bottleneck, this can leave insufficient capacity for the linguistic content needed for accurate reconstruction, thereby hurting intelligibility.

Hybrid approaches incorporate semantic features from self-supervised learning(SSL) models \cite{chen2022wavlm,hsu2021hubert,baevski2020wav2vec} to strengthen semantic integrity and downstream audio language model capabilities \cite{ye2025codec,socodec,jiang2024universal}. However, the semantic features are often not explicitly constrained to focus on linguistic content, and they may contain speaker identity and other paralinguistic attributes \cite{qiang2025secousticodec}. Moreover, in many hybrid designs, semantic cues are used primarily as initial priors; their influence can attenuate across successive decoding stages, leaving the final reconstruction still largely dominated by acoustic fidelity and vulnerable to intelligibility degradation at ultra-low bitrates \cite{defossez2024moshi,zhangspeechtokenizer}.

\begin{figure*}[t]
    \centering
    \includegraphics[width=\textwidth]{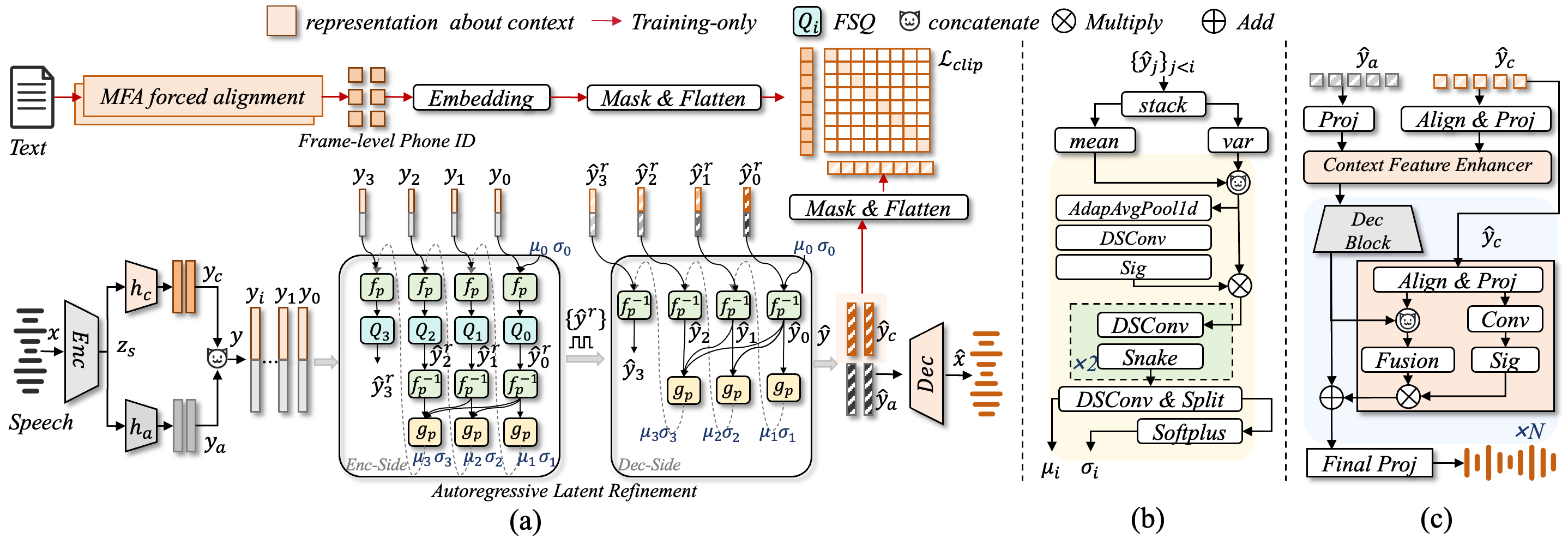}
    \caption{Overview of ContextCodec. (a) The end-to-end pipeline. (b) The predictor $g_p$. (c) The context-guided attention decoder.}
    \label{fig:system_overview}
    \vspace{-14pt}
\end{figure*}

These limitations highlight the need for a content-first codec design at ultra-low bitrates, where the linguistic message must be prioritized over non-linguistic acoustic variation. Motivated by this, we propose ContextCodec, a context-guided neural speech codec for ultra-low bitrate speech communication (Figure.~\ref{fig:introduction}(c)). It adopts a dual-branch encoder that decouples acoustic-detail modeling from content-focused context modeling, and learns linguistically aligned context representations~\cite{radford2021learning} to guide speech reconstruction. The learned context information is then used to support more reliable decoding under severe bitrate.  We then inject the learned context features into each decoding stage to actively guide waveform reconstruction for improved intelligibility. Finally, inspired by learned image codecs~\cite{minnen2018joint,cheng2020learned,he2022channelwise}, we introduce a lightweight autoregressive (AR) latent refinement module that progressively refines latents before decoding, enabling improved reconstruction quality. Our contributions are as follows:

1) We propose ContextCodec, a content-first neural speech codec designed for ultra-low-bitrate communication, which explicitly prioritizes linguistic content through decoupled acoustic and context modeling and a CLIP (Contrastive Language--Image Pretraining)-style linguistic alignment objective.

2) We design a context-guided attention decoder that preserves contextual guidance under tight bitrate constraints through acoustic pre-conditioning and stage-wise context injection, complemented by a lightweight autoregressive latent refinement module that improves reconstruction quality before waveform generation.

3) Extensive objective and subjective evaluations demonstrate that ContextCodec achieves a favorable quality–intelligibility trade-off down to 500 bps with a real-time factor (RTF) of 0.4886 on a typical mobile CPU.

\vspace{-7pt}
\section{Design of ContextCodec}

ContextCodec is built upon a GAN-based quantized autoencoder framework with finite scalar quantization (FSQ) \cite{mentzerfinite}. The base blocks of the shared encoder, decoder, and multiple discriminators follow the same architectural settings as DAC~\cite{dac}.

\vspace{-7pt}
\subsection{Overall architecture}
\label{sec:overall_arch}

As shown in Figure~\ref{fig:system_overview}(a), ContextCodec maps an input waveform $x$ to shared features $z_s=\mathrm{Enc}(x)$ and applies a dual-branch encoder to obtain an acoustic stream $y_a\in\mathbb{R}^{B\times d_a\times T}$ and a context stream $y_c\in\mathbb{R}^{B\times d_c\times T}$; the concatenated latents are denoted by $y=[y_a;y_c]$, where $B$ is batch size and $T$ is the number of codec frames. The overall training pipeline includes three components. \textbf{(1) Autoregressive latent refinement:} $y$ is partitioned into interleaved phases $\{y_i\}_{i=0}^{P-1}$, and a reusable predictor $g_p$ estimates per-phase statistics for phase-specific quantization, yielding refined symbols $\hat{y}^{r}$ for transmission; on the decoder-side, the restored refined latents $\hat{y}$ are reconstructed from $\hat{y}^{r}$. \textbf{(2) CLIP-style phoneme alignment:} $\hat{y}$ is split into $\hat{y}_a$ and $\hat{y}_c$, and the quantized context codes are supervised by a frame-level contrastive loss against forced-aligned phoneme indices from text data. \textbf{(3) Context-guided attention decoding:} the decoder injects $\hat{y}_c$ into each decoding stage to reconstruct the waveform $\hat{x}=\mathrm{Dec}(\hat{y}_a,\hat{y}_c)$.

\vspace{-7pt}
\subsection{Autoregressive latent refinement}
\label{sec:ar_refinement}

To better allocate bits under tight budgets, we refine the concatenated latents $y=[y_a;y_c]$ before decoding by sequentially predicting per-phase mean/scale from already restored phases and quantizing normalized residuals, yielding refined symbols $\hat{y}^{r}$ (and restored $\hat{y}$) in a causal, lightweight manner. Specifically, we split $y$ into $P$ \emph{interleaved} phase sequences $\{y_i\}_{i=0}^{P-1}$ (frames $t\equiv i\ \mathrm{mod}\ P$) and process phases in order ($0\rightarrow P{-}1$) within each $P$-frame block; we describe Enc-side bitstream generation and Dec-side reconstruction separately.

\textbf{Enc-side.} We denote the restored refined result of phase $i$ by $\hat{y}_i$. For $i=0$, we directly quantize with a phase-specific quantizer $Q_0$: $\hat{y}^{r}_{0}=Q_0(y_0)$ and $\hat{y}_0=\hat{y}^{r}_{0}$(set $\mu_0=0,\sigma_0=1$). For each subsequent $i=1,\dots,P{-}1$, we use a reusable predictor $g_p$ to estimate per-time mean and scale from the already restored sub-sequences, $(\mu_i,\sigma_i)=g_p(\{\hat{y}_j\}_{j<i})$. As shown in Figure~\ref{fig:system_overview}(b), the module $g_p$ aggregates the available context by computing per-time mean and standard deviation across $\{\hat{y}_j\}_{j<i}$, applies adaptive average pooling followed by a small 1D convolution and sigmoid as a lightweight global gate, and then uses a small depthwise-separable 1D conv stack with Snake activations to output $(\mu_i,\sigma_i)$. We then remove the predicted mean/scale to form a normalized residual:
\vspace{-5pt}
\begin{equation}
\varepsilon_i = f_p(y_i;\mu_i,\sigma_i) = (y_i-\mu_i)\oslash \sigma_i,
\end{equation}

\vspace{-6pt}
We quantize $\varepsilon_i$ with a phase-specific quantizer and restore it as:
\vspace{-5pt}
\begin{equation}
\hat{y}^{r}_{i}=Q_i(\varepsilon_i),\quad
\hat{y}_i = f_p^{-1}(\hat{y}^{r}_{i};\mu_i,\sigma_i)=\mu_i+\sigma_i\odot \hat{y}^{r}_{i}.
\end{equation}
where $\oslash$ and $\odot$ denote element-wise division and multiplication. The refined quantized sequence $\hat{y}^{r}$ (obtained by interleaving $\lbrace \hat{y}^{r}_{i}\rbrace_{i=0}^{P-1}$ back to the full rate) is then coded into a bitstream and transmitted to the decoder.

\textbf{Dec-side.} The decoder parses the received bitstream to recover the discrete symbols $\hat{y}^{r}$. It then reconstructs the restored refined latents $\hat{y}$ by repeating the same phase order ($0\rightarrow P{-}1$): for $i=0$, set $\hat{y}_0=\hat{y}^{r}_{0}$; for $i\ge 1$, recompute $(\mu_i,\sigma_i)=g_p(\{\hat{y}_j\}_{j<i})$ from already restored phases and apply $\hat{y}_i=f_p^{-1}(\hat{y}^{r}_{i};\mu_i,\sigma_i)$. Finally, interleaving $\lbrace \hat{y}_i\rbrace_{i=0}^{P-1}$ yields the full-rate $\hat{y}$, which is fed into the waveform decoder.

In a future streaming, stateful implementation, phases can be reconstructed in order by caching $\{\hat{y}_j\}_{j<i}$, requiring no extra fixed $P$-frame delay; the added algorithmic latency can remain one codec frame.

\vspace{-7pt}
\subsection{CLIP-style phoneme alignment}
\label{sec:clip_alignment}

 Inspired by Secousticodec~\cite{qiang2025secousticodec}, we introduce a frame-level CLIP-style alignment objective that directly matches the codec context features with the paired phoneme sequence at the codec frame rate. We obtain frame-level phoneme IDs by Montreal Forced Aligner (MFA) forced alignment and map them to learnable vectors by an embedding table as illustrated in Figure~\ref{fig:system_overview}(a). This objective is designed to encourage content-focused, frame-consistent context representations and to reduce speaker-related and other paralinguistic leakage in the context branch. Let $\hat{y}_c\in\mathbb{R}^{B\times d_c\times T}$ denote the frame-wise \emph{quantized} context representation split from $\hat{y}$, and let $q\in\mathbb{N}^{B\times T}$ be the aligned phoneme indices. The embedding table maps $q$ to $e\in\mathbb{R}^{B\times T\times d_c}$. We define a binary mask $m\in\{0,1\}^{B\times T}$ to select valid frames and perform Mask\&Flatten by dropping invalid frames and flattening only valid frame pairs into $N$ samples. The CLIP-style contrastive objective is then computed as:

\vspace{-10pt}
\begin{align}
\ell_{ij} &= \tfrac{\operatorname{sim}(\tilde{y}_{c,i},\tilde{e}_j)}{\tau},\quad i,j=1,\dots,N, \\
\mathcal{L}_{\text{clip}} &= \tfrac{1}{2}\Big(\operatorname{CE}(\ell, t) + \operatorname{CE}(\ell^{\top}, t)\Big),\ \ t_i=i,
\end{align}
where $\tilde{y}_{c},\tilde{e}$ are $\ell_2$-normalized, $\operatorname{sim}(\cdot,\cdot)$ is cosine similarity, and $\tau$ is temperature. The positive pair of sample $i$ is the aligned (utterance, frame) pair $(\tilde{y}_{c,i},\tilde{e}_i)$, while negatives are all other valid frames in the mini-batch.

\begin{table*}[t]
\centering
\caption{Objective evaluation results on a multilingual set and VCTK. Best results are shown in bold.}
\vspace{-5pt}
\renewcommand{\arraystretch}{0.99}
\setlength{\tabcolsep}{2pt}
\begin{tabular}{@{}l|c|c|c|c|cccc|cccc@{}}
\toprule
\multirow{2}{*}{Model} & \multirow{2}{*}{Type} & \multirow{2}{*}{Bitrate} & \multirow{2}{*}{\begin{tabular}{c}Param\\(M)\end{tabular}} & \multirow{2}{*}{\begin{tabular}{c}Sample\\rate\end{tabular}} & \multicolumn{4}{c|}{Multilingual (10 langs)} & \multicolumn{4}{c}{VCTK (en)} \\
\cmidrule(lr){6-9}\cmidrule(lr){10-13}
 & & & & & PESQ$\uparrow$ & STOI$\uparrow$ & SI-SDR$\uparrow$ & WER$\downarrow$ & PESQ$\uparrow$ & STOI$\uparrow$ & SI-SDR$\uparrow$ & WER$\downarrow$ \\
\midrule
\multicolumn{13}{c}{\textbf{around 1000 bps Comparison}} \\
\midrule
EnCodec~\cite{encodec} & Acoustic & 1500 & \textbf{14.85} & 24~kHz & 1.629 & 0.814 & 0.511 & 45.76\% & 1.786 & 0.796 & 0.777 & 9.60\% \\
DAC~\cite{dac} & Acoustic & 2000 & 74.18 & 16~kHz & 1.201 & 0.727 & -11.940 & 77.98\% & 1.245 & 0.740 & -14.125 & 20.84\% \\
SNAC~\cite{siuzdak2024snac} & Acoustic & 994 & 19.80 & 24~kHz & 1.835 & 0.834 & -4.119 & 39.32\% & 2.384 & 0.861 & 0.115 & 5.28\% \\
Secousticodec~\cite{qiang2025secousticodec} & Hybrid & 1090 & 94.75 & 22~kHz & 1.600 & 0.804 & -26.675 & 56.14\% & 1.880 & 0.847 & -25.572 & 9.40\% \\
SemantiCodec~\cite{semanticodec} & Hybrid & 1250 & 699.41 & 16~kHz & 1.882 & 0.828 & -25.758 & 30.95\% & 2.103 & 0.848 & -23.693 & 4.62\% \\
FACodec~\cite{ju2024naturalspeech} & Hybrid & 1600 & 102.33 & 16~kHz & 1.310 & 0.743 & -24.937 & 51.79\% & 1.623 & 0.788 & -14.983 & 4.15\% \\
SpeechTokenizer~\cite{zhangspeechtokenizer} & Hybrid & 1000 & 103.68 & 16~kHz & 1.242 & 0.715 & -8.789 & 83.12\% & 1.351 & 0.747 & -6.449 & 8.40\% \\
X-Codec~\cite{ye2025codec} & Hybrid & 1000 & 169.33 & 16~kHz & 1.846 & 0.812 & -18.693 & 31.06\% & 2.257 & 0.822 & -17.611 & 3.29\% \\
Mimi~\cite{defossez2024moshi} & Hybrid & 1100 & 79.31 & 24~kHz & 2.028 & 0.852 & 1.614 & 33.60\% & 2.256 & 0.840 & 2.968 & 4.57\% \\
ContextCodec (ours) & Hybrid & 1000 & 78.61 & 16~kHz & \textbf{2.140} & \textbf{0.866} & \textbf{2.110} & \textbf{28.31\%} & \textbf{2.476} & \textbf{0.880} & \textbf{3.614} & \textbf{2.25\%}  \\
\midrule
\multicolumn{13}{c}{\textbf{around 500 bps Comparison}} \\
\midrule
Mimi~\cite{defossez2024moshi} & Hybrid & 550 & 79.31 & 24~kHz & 1.553 & 0.790 & -5.517 & 60.35\% & 1.685 & 0.772 & -4.285 & 10.22\% \\
SpeechTokenizer~\cite{zhangspeechtokenizer} & Hybrid & 500 & 103.68 & 16~kHz & 1.150 & 0.590 & -37.434 & 107.42\% & 1.210 & 0.645 & -37.209 & 10.53\% \\
SemantiCodec~\cite{semanticodec} & Hybrid & 625 & 699.41 & 16~kHz & 1.660 & 0.788 & -26.761 & 52.69\% & 1.910 & 0.820 & -24.128 & 11.42\% \\
ContextCodec (ours) & Hybrid & 500 & \textbf{78.61} & 16~kHz & \textbf{1.758} & \textbf{0.812} & \textbf{-2.648} & \textbf{52.11\%} & \textbf{2.120} & \textbf{0.846} & \textbf{-0.604} & \textbf{5.85\%} \\
\bottomrule
\end{tabular}
\renewcommand{\arraystretch}{1}
\label{tab:results}
\vspace{-5pt}
\end{table*}

\vspace{-7pt}
\subsection{Context-guided attention decoder}
\label{sec:context_decoder}

As shown in Figure~\ref{fig:system_overview}(c), the context-guided attention decoder $Dec(\cdot)$ injects the quantized context representation $\hat{y}_c$ into waveform reconstruction, so that the context branch can actively guide generation. Rather than conditioning the decoder only at its input, we use context features to pre-condition the acoustic latents and to modulate each subsequent decoding stage, which helps preserve contextual guidance when reconstructing from heavily quantized representations.

Given the restored refined latents $\hat{y}=f_p^{-1}(\hat{y}^{r})$, we split them along the channel dimension into an acoustic stream $\hat{y}_a$ and a context stream $\hat{y}_c$. Before upsampling, we apply a lightweight \emph{context feature enhancer} to modulate $\hat{y}_a$ using $\hat{y}_c$ through two complementary pathways: (i) a global pathway that produces a channel-wise gate and (ii) a local pathway that predicts a time-varying residual. The gated and residual-modulated features are concatenated and fused by a small fusion network to produce context-conditioned acoustic features.    The two pathways and the fusion network are realized with lightweight pointwise and kernel-3 1D convolutions with Snake activations.

We then progressively upsample the acoustic stream with transposed-convolution decoder blocks and residual 1D convolution units with Snake activations. At each decoding stage, the context stream is aligned to the current temporal resolution by linear interpolation and projected to the acoustic channel size with a pointwise convolution, which corresponds to \emph{Align\&Proj} in Figure~\ref{fig:system_overview}(c). The projected context features also produce a sigmoid gate, denoted as \emph{Sig}, and the stage-wise \emph{Fusion} concatenates acoustic and projected context features and applies the same kernel-3 convolution, Snake, and pointwise convolution. The fused output is gated and added back to the acoustic features via a residual connection. The final waveform is produced by Snake, a kernel-7 1D convolution, and a $tanh$ output layer.

\vspace{-7pt}
\subsection{Training objectives}

We jointly optimize reconstruction, adversarial, and cross-modal alignment objectives. The generator loss is:
\vspace{-2pt}
\begin{align}
\mathcal{L}_{G} = \lambda_{m}\mathcal{L}_{\text{m}} + \lambda_{\text{adv}}\mathcal{L}^{G}_{\text{adv}} + \lambda_{\text{fm}}\mathcal{L}_{\text{fm}} + \lambda_{\text{clip}}\mathcal{L}_{\text{clip}},
\end{align}
where $\mathcal{L}_{\text{m}}$ is the multi-scale mel loss, $\mathcal{L}^{G}_{\text{adv}}$ and $\mathcal{L}_{\text{fm}}$ are the GAN generator and feature-matching losses. $\mathcal{L}_{\text{clip}}$ is a symmetric InfoNCE loss. The discriminator is optimized with $\mathcal{L}_{D}{=}\mathcal{L}^{D}_{\text{adv}}$.

\vspace{-4pt}
\section{Experiments}
\vspace{-2pt}

\subsection{Experimental setup}
\vspace{-2pt}

\textbf{Datasets and preprocessing.} ContextCodec is trained on the training sets of LibriTTS~\cite{zen2019libritts} and AISHELL-3~\cite{shi2020aishell}. To obtain frame-level supervision for alignment, we run an offline forced-alignment pipeline using MFA \footnote{\url{https://montreal-forced-aligner.readthedocs.io/}} to produce phoneme-level TextGrids~\cite{mahrt2016praatio}. We then convert phoneme time intervals into a per-frame phoneme-id sequence at the codec frame rate. All training audio is segmented into 3-second clips. For evaluation, we test on VCTK~\cite{yamagishi2012english} and ten languages from Common Voice 21.0~\cite{commonvoice:2020}, including English, Chinese, German, French, Spanish, Russian, Arabic, Hindi, Japanese, and Korean to validate reconstruction performance and cross-lingual and cross-domain generalization. We randomly sample 6{,}000 VCTK utterances and 300 utterances per CV21 language. Each baseline runs inference at its native sampling rate, while all objective metrics are computed at 16 kHz for fair comparison.

\textbf{Model and training.} Training is performed at 16 kHz. The default model uses $N_e{=}4$ downsampling blocks with rates $[4,4,5,8]$ and hop size $h{=}\prod r_i{=}640$. Both heads output $d_a{=}d_c{=}512$ channels; the context head uses $N_T{=}2$ Transformer layers with 4 attention heads. We choose $P{=}4$ phases and each phase has an independent FSQ quantizer with unequal vector. For 500~bps we use $\{[8\times6,4\times2],[8\times6,4\times1],[8\times6,4\times1],[8\times6]\}$ and for 1000~bps we increase capacity to $\{[8\times12,4\times4],[8\times12,4\times2],[8\times12,4\times2],[8\times12]\}$. Optimization uses AdamW with lr $2\times 10^{-4}$ and betas $(0.8,0.99)$, and we use GAN training with the same discriminators as DAC~\cite{dac}. The temperature $\tau{=}0.07$ for CLIP-style alignment. We train for 1M steps on a single NVIDIA RTX 4090 GPU with batch size 8, using $\lambda_m{=}15$, $\lambda_{\text{adv}}{=}1$, $\lambda_{\text{fm}}{=}2$, $\lambda_{\text{clip}}{=}3$.

\textbf{Baselines and metrics.} We compare against representative baselines summarized in Table~\ref{tab:results} using their official pretrained weights and recommended inference settings. For objective evaluation, we report signal-level metrics and the word error rate (WER) computed by a pre-trained Whisper-Turbo~\cite{whisper}. For each test language, we explicitly set Whisper-Turbo to the corresponding language when computing WER. For subjective evaluation, we conduct a preliminary pairwise preference listening test~\cite{schoeffler2018webmushra} with 15 listeners on 15 VCTK utterances sampled by stratified random selection, and all test outputs are loudness-normalized to -16 dB before playback.

\vspace{-6pt}
\subsection{Results}
\vspace{-4pt}

\begin{table}[t]
\vspace{-5pt}
\centering
\caption{Subjective pairwise preference test results (\%)}
\vspace{-8pt}
\small
\renewcommand{\arraystretch}{0.95}
\setlength{\tabcolsep}{3pt}
\begin{tabular}{@{}lclc@{}}
\toprule
codec & preference (\%) & codec & No pref. (\%) \\
\midrule
ContextCodec & 97.92~$\leftarrow$~2.08 & Opus 6K~\cite{opus} & 0.00 \\
ContextCodec & 29.17~$\rightarrow$~54.17 & Reference & 16.66 \\
ContextCodec & 52.92~$\leftarrow$~40.83 & SemantiCodec & 6.25 \\
Reference & 66.67~$\leftarrow$~20.83 & SemantiCodec & 12.50 \\
\bottomrule
\end{tabular}
\renewcommand{\arraystretch}{1}
\vspace{-8pt}
\label{tab:pref_test}
\end{table}

\textbf{Objective results.} Table~\ref{tab:results} summarizes objective results. Overall, ContextCodec achieves strong performance on both English and multilingual test sets, offering a favorable trade-off between intelligibility and perceptual quality at ultra-low bitrates with a compact model size. These results also suggest reasonable generalization to unseen Common Voice languages, likely benefited by our phoneme-index supervision, which provides a shared pronunciation-based signal across different scripts. Nevertheless, phonological inventories differ across languages, and rare or unseen phonemes may be underrepresented in training.

\textbf{Subjective results.} We further conduct a subjective pairwise preference test as shown in Table~\ref{tab:pref_test}, comparing against SemantiCodec, the strongest hybrid baseline in our low-bitrate objective comparison, and Opus as a widely used traditional codec: at 500~bps, ContextCodec is preferred over both baselines, indicating strong perceptual quality at ultra-low bitrates.

\textbf{Deployment efficiency.} Computational complexity\footnote{GMACs is the maximum across the ptflops, fvcore, and thop profilers. RTF is averaged over 10 runs on 10-second speech after warm-up.} is evaluated in Table~\ref{tab:complexity}. For on-device evaluation, we report the end-to-end RTF on a Snapdragon 8 Gen 3 phone using CPU only and standard precision. Results show that ContextCodec remains efficient enough for practical deployment.

\begin{table}[t]
\centering
\vspace{0pt}
\caption{Complexity and Efficiency Comparison.}
\vspace{-8pt}
\small
\renewcommand{\arraystretch}{0.95}
\setlength{\tabcolsep}{4pt}
\begin{tabular*}{\columnwidth}{@{\extracolsep{\fill}}lccc}
\toprule
Model & GMACs $\downarrow$ & \begin{tabular}[c]{@{}c@{}}RTF\\(A100)$\downarrow$\end{tabular} & \begin{tabular}[c]{@{}c@{}}RTF\\(Android)$\downarrow$\end{tabular} \\
\midrule
ContextCodec & 22.55 & 0.0029 & \textbf{0.4886} \\
X-Codec & 30.80 & \textbf{0.0027} & - \\
Mimi & \textbf{11.61} & 0.0049 & - \\
\bottomrule
\end{tabular*}
\renewcommand{\arraystretch}{1}
\vspace{-20pt}
\label{tab:complexity}
\end{table}

\vspace{-8pt}
\subsection{Ablation and analysis}
\vspace{-4pt}

\begin{table}[t]
\vspace{-4pt}
\caption{Ablation study on LibriTTS test-clean. ``Enh.'' denotes stage-wise context injection.}
\vspace{-10pt}
\label{tab:ablation_final}
\centering
\small
\renewcommand{\arraystretch}{0.95}
\setlength{\tabcolsep}{2pt}
\begin{tabular*}{\columnwidth}{@{\extracolsep{\fill}}c>{\raggedright\arraybackslash}p{2.2cm}cccccc@{}}
\toprule
ID & Supervision & $\lambda_{\text{clip}}$ & Enh. & $P$ & PESQ$\uparrow$ & STOI$\uparrow$ & WER$\downarrow$ \\ \midrule
$\mathrm{M}_{0}$ & Phoneme-CLIP & 3.0 & \checkmark & 4 & 2.048 & 0.892 & \textbf{5.56\%} \\ \midrule
$\mathrm{M}_{1}$ &  SSL-Distill\cite{baevski2020wav2vec} & -- & \checkmark & 4 & 2.079 & 0.895 & 7.91\% \\
$\mathrm{M}_{2}$ &  Phoneme-CLIP & 0.5 & \checkmark & 4 & \textbf{2.114} & \textbf{0.900} & 9.14\% \\
$\mathrm{M}_{3}$ &  None & 0 & \checkmark & 4 & 2.100 & 0.899 & 10.58\% \\ \midrule
$\mathrm{M}_{4}$ &  Phoneme-CLIP & 3.0 & $\times$ & 4 & 2.080 & 0.896 & 8.20\% \\ \midrule
$\mathrm{M}_{5}$ &  None & 0.0 & $\times$ & 4 & 2.047 & 0.893 & 10.75\% \\
$\mathrm{M}_{6}$ &  None & 0.0 & $\times$ & 2 & 1.963 & 0.884 & 10.51\% \\
$\mathrm{M}_{7}$ &  None & 0.0 & $\times$ & 0 & 1.887 & 0.880 & 10.75\% \\ \bottomrule
\end{tabular*}
\vspace{-2pt}
\end{table}

As Table~\ref{tab:ablation_final} shows, we investigate the contribution of each proposed component by comparing against several variants and denote the variants from top to bottom as $\mathrm{M}_{0}$--$\mathrm{M}_{7}$: (i) $\mathrm{M}_{1}$ replaces CLIP-style alignment with knowledge distillation from pre-trained Wav2Vec 2.0~\cite{baevski2020wav2vec}. We extract teacher representations by averaging the hidden states from \emph{all} hidden layers, align them to codec frames, and match them to the student context features $\hat{y}_c$ with a cosine-similarity distillation loss weighted by $\lambda_{\text{distill}}{=}3$; (ii) $\mathrm{M}_{2}$ and $\mathrm{M}_{3}$ reduce the CLIP loss weight to 0.5 and 0; (iii) $\mathrm{M}_{4}$ adopts the default decoder~\cite{dac} with stage-wise context injection disabled; (iv) $\mathrm{M}_{5}$, $\mathrm{M}_{6}$, and $\mathrm{M}_{7}$ vary the number of AR refinement phases $P$ to 4, 2, and 0 (no refinement). It supports three conclusions. First, CLIP-style phoneme alignment provides more effective semantic supervision for intelligibility than distillation ($\mathrm{M}_{0}$ vs. $\mathrm{M}_{1}$). Second, explicit context guidance is beneficial, and the proposed stage-wise context injection mechanism is more effective than decoding from concatenated latents without guidance ($\mathrm{M}_{0}$ vs. $\mathrm{M}_{4}$). Third, AR refinement improves perceptual quality ($\mathrm{M}_{5}$ vs. $\mathrm{M}_{7}$).

\begin{table}[t]
\centering
\vspace{-7pt}
\caption{Attribute predictability analysis (accuracy, \%).}
\vspace{-4pt}
\small
\renewcommand{\arraystretch}{0.95}
\setlength{\tabcolsep}{4pt}
\begin{tabular*}{\columnwidth}{@{\extracolsep{\fill}}clccc}
\toprule
ID & Supervision & Phone $\uparrow$ & Speaker $\downarrow$ & Dialect $\downarrow$ \\
\midrule
$\mathrm{M}_{0}$ & Phoneme-CLIP & \textbf{88.7} & \textbf{51.8} & \textbf{26.6} \\
$\mathrm{M}_{1}$ & SSL-Distill & 70.2 & 91.0 & 28.2 \\
$\mathrm{M}_{3}$ & None & 66.5 & 82.2 & 30.8 \\
\bottomrule
\end{tabular*}
\renewcommand{\arraystretch}{1}
\vspace{-18pt}
\label{tab:rep_purity_timit}
\end{table}

\textbf{Attribute predictability analysis.} We quantify attribute predictability on TIMIT using linear probes for three supervision settings $\mathrm{M}_{0}$/$\mathrm{M}_{1}$/$\mathrm{M}_{3}$ in Table~\ref{tab:ablation_final}. For each setting, we form two types of embeddings from the frame-level quantized context representation $\hat{y}_c$ for probing: an utterance-level embedding via time-averaging, and a phoneme-segment-level embedding using TIMIT boundaries (specifically for the Phone probe). We then train logistic-regression probes to predict phone, speaker, and dialect. Phone and Dialect probes are trained on the TRAIN split and evaluated on the TEST split. For Speaker, since TRAIN and TEST contain disjoint speakers, we instead split each TRAIN speaker's utterances into train and test sets and report accuracy on the test sets. As shown in Table~\ref{tab:rep_purity_timit}, Phoneme-CLIP substantially improves Phone accuracy and reduces the predictability of Dialect and Speaker compared with SSL-Distill, indicating more content-specific representations with lower speaker/dialect leakage.

\vspace{-9pt}
\section{Conclusion}
\vspace{-4pt}

We propose ContextCodec, a context-guided neural speech codec for ultra-low bitrate speech communication. By prioritizing content preservation and using context as an explicit guidance signal during decoding, ContextCodec achieves a strong balance between intelligibility and perceptual quality at bitrates down to 500~bps. Extensive objective and subjective evaluations further demonstrate robust performance across English and multilingual test sets, together with an RTF of 0.4886 on a typical mobile CPU. Future work includes the streaming implementation and strengthening multilingual supervision for rare or unseen phonemes.

\ifcameraready
\section{Acknowledgements}

This work is supported by the National Key Research and Development Program of China under Grant No. 2023YFB2904300, the National Natural Science Foundation of China under Grant No. 62293484, and Beijing Natural Science Foundation (F251001).
\fi

\section{Generative AI Use Disclosure}
We used generative AI tools (GPT-5.2) to assist with English polishing and \LaTeX{} formatting suggestions. All technical content, experimental results, and final wording were verified and approved by the authors.

\bibliographystyle{IEEEtran}
\bibliography{refs_liang}

\end{document}